\documentclass{revtex4}
\usepackage{epsf,graphicx,subfigure,amsmath,amssymb,amsfonts}

\begin{document}
\paperwidth =15cm 
%\twocolumn
% \draft command makes pacs numbers print
%\tighten
\title{Comment on: ``Nonextensivity: from low-dimensional maps to
  Hamiltonian systems'' by Tsallis et al.\cite{tsallis02}}
\author{D.H.E. Gross}
\address{
Hahn-Meitner-Institut
Berlin, Bereich Theoretische Physik,Glienickerstr.100\\ 14109 Berlin, Germany
and 
Freie Universit{\"a}t Berlin, Fachbereich Physik; \today}
\begin{abstract} 
  The critique against using Boltzmann's {\em microcanonical} entropy,
  an "ensemble measure", as foundation of statistics is rebuffed.  The
  confusion of the microcanonical distribution with the exponential
  Boltzmann-Gibbs (``BG'') distribution is pointed out.  Boltzmann's
  principle is clearly superior over any Tsallis q-statistics in
  describing the equilibrium of small systems like nuclei and even
  self-gravitating systems as paradigm of non-extensive Hamiltonian
  systems.
\end{abstract}
\maketitle Since 1981 thermo-statistics is addressed to highly excited
(``hot'') nuclei~\cite{gross45,gross95} and a little bit later to
atomic clusters\cite{gross154,gross151,gross152,gross157,gross174}.
Not only the mass-dispersion, fragment-mass fluctuations, but most
importantly the negative heat-capacity at fragmentation was predicted:
Of course these systems cannot be treated in the thermodynamic limit
(they are ``non-extensive'') and many gospels of traditional canonical
thermo-statistics have to be abolished. E.g. the canonical ensemble
fails, it is not equivalent to the fundamental microcanonical
ensemble, the specific heat can become {\em negative} and Clausius
formulation of the Second Law is violated: ``heat can only flow from
hot to cold'', phase transitions are found unambiguously and sharply
in all details in these small systems~\cite{gross189,gross174}.  This
is the reason why this extension of statistics met severe resistance.
Only recently after more than $20$ years it becomes widely accepted in
the nuclear physics community.  There are meanwhile many experimental
results, c.f.~\cite{borderie02}, earlier ones listed
in~\cite{gross189}, that confirm in great detail this new approach to
the thermo-dynamics of these non-extensive systems including the
``exotic'' features mentioned above, see however~\footnote{It is true
  that Tsallis et al. suggested in one case of an experimentally
  confirmed negative heat capacity (sodium
  clusters~\protect\cite{schmidt01}) an alternative
  explanation~\protect\cite{tsallis01}: In their 1-dim HMF-model they
  found a metastable state with most likely zero microcanonical
  measure, i.e. with no entropy. (It plays no role in the complete
  microcanonical ensemble, where is no sign of a
  backbending~\cite{gross174}.) The average kinetic energy in this
  state shows a back-bending as function of the total energy. Now, the
  microcanonical temperature is $T=(\partial S_B(E)/\partial E)^{-1}$
  and refers evidently to the whole microcanonical ensemble not to a
  single trajectory with zero measure. In view of this and the very
  qualitative nature and flatness of this ``explanation'' which is in
  contrast to the many other, very detailed and specific verifications
  of the prediction by the Boltzmann microcanonical ensemble listed
  above, one can safely ignore this alternative.}.  Do we really need
another $20$ years to accept these ideas in the wider community of
Statistical Physics? It seems so:

The conference held in Febr.2002 in Les Houches on "Dynamics and
Thermodynamics of Systems with Long Range Interactions" addressed
explicitely these non-extensive systems.  In their
contribution~\cite{tsallis02} to this conference Tsallis, Rapisarda,
Latora and Baldovin illuminated the wide range of application of
Tsallis q-statistics, again a {\em canonical}
approach~\cite{tsallis99}. The main object of this formalism is the
{\em dynamics} of many-body systems {\em out of equilibrium}.  E.g.
the change of the distance $\xi(t)$ of two initially neighboring
points (the sensitivity to the initial conditions) under the logistic
map $x_{t+1}=1-ax_t^2$ follows a q-exponential of $t$. $q$ controls
how strong the mixing of the dynamics is, which is a condition for
equilibration.

The q-{\em entropy} addresses the distribution in phase-space: E.g. a
narrow set of points develops under the logistic map with parameter
$a>a_c=1.4101\cdots$, with $S_{q=1}(t)$ rising linearly in time,
whereas with $a=a_c$, i.e. at the edge of chaos
$S_{q=0.2445\cdots}(t)$ is linear in time.

Now to Tsallis' ``Sancta Sanctorum'' of statistical
mechanics\cite{tsallis02a}, the statistical equilibrium of {\em
Hamiltonian} systems.  For a Hamiltonian system, the uniform
population of the microcanonical manifold ${\cal{E}}$ is the
definition of the {\em equilibrium} distribution. Its geometrical
size $e^{S_B}=tr[\delta(E-H)]$ defines the equilibrium entropy,
Boltzmann's entropy $S_B(E)$. Even the $HMF$-model discussed by
Tsallis approaches it for a finite number of particles in the limit
$t\to\infty$~\cite{tsallis02}.  There Tsallis' q-entropy has $q=1$
and is identical to $S_B$. This finding agrees with my
conclusion~\cite{gross185} and also in this book~\cite{gross189} that
equilibrium Hamiltonian systems have $S_B$.  Here it is essential to
realize that Boltzmann's entropy refers to the {\em microcanonical}
uniform population of the energy-manifold ${\cal{E}}$ and {\em not}
to the canonical ``Boltzmann-Gibbs'' (BG) distribution as claimed
by~\cite{tsallis02}. The difference is important at
phase-separations. Here the canonical ensemble is not equivalent to
the micro-ensemble not only for non-extensive systems (what is
trivially the case in general) but also in the thermodynamic limit of
ordinary extensive systems~\cite{gross174,gross189}. Boltzmann's
entropy $S_B(E)$ is well defined, multiply differentiable even at
phase transitions, independently of whether it is extensive or not,
i.e.  $S_B(A+B)$ equals $S_B(A)+S_B(B)$ or not. I.e. the eventual
non-extensivity of Hamiltonian systems does not demand any exotic
entropy at equilibrium.

Before introducing Tsallis' non-extensive, canonical $q$-entropy one
should better exploit the original {\em microcanonical} Boltzmann's
statistics.  Precisely, this is done by my geometric approach to
statistical mechanics. Its success to predict the most sophisticated,
and from the view of conventional canonical statistics exotic and
surprising phenomena of phase transitions in small systems like hot
nuclei was mentioned above. There is yet no alternative theory.

In this context a further remark: Tsallis et al.~\cite{tsallis02}
quote Einstein's objection against the use of $S_B(E)$ in a lengthy
discussion about the additivity (``extensivity'') of $S$ for
independent, non-interacting systems. Einsteins remark has nothing to
do with the additivity of $S_B(E)$ for independent systems. It
concerns the definition of equilibrium values of some macroscopic
observables as time-averages $\overline{A}$ compared to Boltzmann's
second definition as ensemble averages $<\!\!A\!\!>$. The advantage of
the ensemble-probabilistic definition of $S_B(E)$ and of $<\!\!A\!\!>$
for non-extensive or small systems compared to time-averages
$\overline{A}$ as favored by Einstein was in detail discussed already
in~\cite{gross183} see also~\cite{gross180,gross186}.

Up to now the most realistic application of Boltzmann (not
``BG''!)statistics to produce the microcanonical phase diagram
of self-gravitating systems under various angular-momenta is
given in~\cite{gross187}. Without any doubt this system is the paradigm
of non-extensive Hamiltonian systems. Of course the singularity of the
gravitation at high density must be shielded like in the Lynden-Bell
statistics~\cite{lyndenbell67} which we use in~\cite{gross187}. At
these densities the relevant physics is ruled by nuclear processes,
like hydrogen burning, and has nothing to do with gravitation, has a
completely different time scale and has also nothing to do with
q-statistics~\cite{tsallis99}. In the whole Tsallis program there is
by far nothing similarly realistic for non-extensive equilibrium
systems. There is no alternative to the microcanonical Boltzmann
statistics and to our geometrical foundation of equilibrium statistics
applied to self-gravitating and rotating systems~\cite{gross187}.
\bibliographystyle{unsrt}%{alpha}%{plain} %{unsrt}
\bibliography{gross,othbiba,othbibb,othbibcd,othbibe,othbibf,othbibg,othbibh,othbibij,othbibk,othbibl,othbibm,othbibn,othbibo,othbibp,othbibr,othbibs,othbibt,othbibuw,othbibxz}

\begin{thebibliography}{10}

\bibitem{tsallis02}
C.~Tsallis, A.~Rapisarda, V.~Latora, and F.~Baldovin.
\newblock Nonextensivity: from low-dimensional maps to Hamiltonian systems.
\newblock In T.~Dauxois, S.~Ruffo, E.~Arimondo, and M.~Wilkens Eds., editors,
  {\em Lecture Notes in Physics}, number 602, pages 148--171, Heidelberg, 2002.
  Springer.

\bibitem{gross45}
D.H.E. Gross and Meng Ta-chung.
\newblock Production mechanism of large fragments in high energy nuclear
  reactions.
\newblock In {\em 4th Nordic Meeting on Intermediate and High Energy Physics},
  page~29, Geilo Sportell, Norway, January 1981.

\bibitem{gross95}
D.H.E. Gross.
\newblock Statistical decay of very hot nuclei, the production of large
  clusters.
\newblock {\em Rep.Progr.Phys.}, 53:605--658, 1990.

\bibitem{gross154}
D.H.E. Gross, M.E. Madjet, and O.~Schapiro.
\newblock Fragmentation phase transition in atomic clusters I ---
  microcanonical thermodynamics.
\newblock {\em Z.Phys.D}, 39:75--83,
  1997;http://xxx.lanl.gov/abs/cond-mat/9608103.

\bibitem{gross151}
M.E. Madjet, D.H.E. Gross, P.A. Hervieux, and O.~Schapiro.
\newblock Fragmentation phase transition in atomic clusters II --- symmetry of
  Coulombic fission.
\newblock {\em Z.Physik D}, 39:309--316,
  1997;http://xxx.lanl.gov/abs/cond-mat/9610118.

\bibitem{gross152}
O.~Schapiro, P.J. Kuntz, K.~M\"ohring, P.A. Hervieux, D.H.E. Gross, and M.E.
  Madjet.
\newblock Fragmentation phase transition in atomic clusters III --- Coulomb
  explosion of metal clusters.
\newblock {\em Z.Physik D}, 41:219--227,
  1997;http://xxx.lanl.gov/abs/cond-mat/9702183.

\bibitem{gross157}
D.H.E. Gross and M.E. Madjet.
\newblock Fragmentation phase transition in atomic clusters IV --- the relation
  of the fragmentation phase transition to the bulk liquid-gas transition.
\newblock {\em Z.Physik B}, 104:541--551, 1997; and
  http://xxx.lanl.gov/abs/cond-mat/9707100.

\bibitem{gross174}
D.H.E. Gross.
\newblock {\em Microcanonical thermodynamics: Phase transitions in ``Small''
  systems}, volume~66 of {\em Lecture Notes in Physics}.
\newblock World Scientific, Singapore, 2001.

\bibitem{gross189}
D.H.E. Gross.
\newblock Thermo-statistics or topology of the microcanonical entropy surface.
\newblock In T.Dauxois, S.Ruffo, E.Arimondo, and M.Wilkens, editors, {\em
  Dynamics and Thermodynamics of Systems with Long Range Interactions}, Lecture
  Notes in Physics, pages 21--45,cond--mat/0206341, Heidelberg, 2002. Springer.

\bibitem{borderie02}
B.~Borderie.
\newblock Dynamics and thermodynamics of the liquid-gas phase transition in hot
  nuclei studied with the Indra array.
\newblock {\em J.Phys.G: Nucl.Part.Phys.}, 28:R212--R227, 2002.

\bibitem{tsallis99}
C.~Tsallis.
\newblock Nonextensive statistics: Theoretical, experimental and computational
  evidences and connections.
\newblock {\em Braz.Journ.Physics}, 29:1, 1999.

\bibitem{tsallis02a}
C.~Tsallis.
\newblock Entropic nonextensivity: a possible measure of complexity.
\newblock {\em Chaos,Solitons,and Fractas}, page~13, 2002.

\bibitem{gross185}
D.H.E. Gross.
\newblock Non-extensive Hamiltonian systems follow Boltzmann's principle not
  Tsallis statistics. -- phase transitions, second law of thermodynamics.
\newblock {\em Physica A}, 305:99--105, cond--mat/0106496, 2002.

\bibitem{gross183}
D.H.E. Gross.
\newblock Ensemble probabilistic equilibrium and non-equilibrium thermodynamics
  without the thermodynamic limit.
\newblock In Andrei Khrennikov, editor, {\em Foundations of Probability and
  Physics}, number XIII in PQ-QP: Quantum Probability, White Noise Analysis,
  pages 131--146, Boston, October 2001. ACM, World Scientific.

\bibitem{gross180}
D.H.E. Gross.
\newblock Second law of thermodynamics, macroscopic observables within
  Boltzmann's principle but without thermodynamic limit.
\newblock {\em submitted to Phys.Rev.E}, page~7, 2000;
  http://arXiv.org/abs/cond-mat/0101281.

\bibitem{gross186}
D.H.E. Gross.
\newblock Geometric foundation of thermo-statistics, phase transitions, second
  law of thermodynamics, but without thermodynamic limit.
\newblock {\em PCCP}, 4:863--872,http://arXiv.org/abs/cond--mat/0201235,
  (2002).

\bibitem{gross187}
E.V. Votyakov, H.I. Hidmi, A.~De Martino, and D.H.E. Gross.
\newblock Microcanonical mean-field thermodynamics of self-gravitating and
  rotating systems.
\newblock {\em Phys.Rev.Lett.}, 89:031101--1--4;
  http://arXiv.org/abs/cond--mat/0202140, (2002).

\bibitem{lyndenbell67}
D.~Lynden-Bell.
\newblock Statistitcal mechanics of violent relaxation in stellar systems.
\newblock {\em Mon. Not. R. astr. Soc.}, 136:101--121, 1967.

\bibitem{schmidt01}
M.~Schmidt, R.~Kusche, T.~Hippler, J.~Donges, W.~Kornm\"uller, B.~von
  Issendorff, and H.~Haberland.
\newblock Negative heat capacity for a cluster of 147 sodium stoms.
\newblock {\em Phys.Rev.Lett.}, 86:1191--1194, 2001.

\bibitem{tsallis01}
C.~Tsallis, B.J.C. Cabral, A.~Rapisarda, and V.~Latora.
\newblock Comment on "negative specific heat for a cluster of 147 sodium atoms"
  by Schmidt et al.
\newblock pages cond--mat/0112266, 2001.

\end{thebibliography}
\end{document}